\documentclass[prl,letterpaper,aps,floatfix,twocolumn]{revtex4-1}
\usepackage{graphicx}
\usepackage{amsmath}
\usepackage{subfigure}
\newcommand{\beq}{\begin{equation}}
\newcommand{\eeq}{\end{equation}}
\newcommand{\bk}{{{\bf{k}}}}

\newcommand{\br}{{{\bf{r}}}}

\newcommand{\bA}{{\bf{A}}}
\newcommand{\bB}{{\bf{B}}}

\newcommand{\bq}{{\bf{q}}}

\newcommand{\beqa}{\begin{eqnarray}}
\newcommand{\eeqa}{\end{eqnarray}}
\newcommand{\pdg}{{\vphantom \dag}}
\newcommand{\dg}{{\dag}}
 
\newcommand{\bsigma}{{\boldsymbol \sigma}}
\newcommand{\btau}{{\boldsymbol \tau}}

\newcommand{\upa}{\uparrow}
\newcommand{\da}{\downarrow} 

\newcommand{\cH}{{\cal H}}
\newcommand{\cI}{{\cal I}}

\newcommand{\cD}{{\cal D}}
\begin{document}
\title{Chiral Anomaly and Diffusive Magnetotransport in Weyl Metals}
\author{A.A. Burkov}
\affiliation{Department of Physics and Astronomy, University of Waterloo, Waterloo, Ontario 
N2L 3G1, Canada} 
\date{\today}
\begin{abstract}
We present a microscopic theory of diffusive magnetotransport in Weyl metals
and clarify its relation to chiral anomaly. We derive coupled diffusion equations for the 
total and axial charge densities and show that chiral anomaly manifests as 
a magnetic-field-induced coupling between them. We demonstrate that 
a universal experimentally-observable consequence of this coupling in magnetotransport 
in Weyl metals is a quadratic negative magnetoresistance, which will dominate all 
other contributions to magnetoresistance under certain conditions.   
\end{abstract}
\maketitle
Weyl semimetals have attracted considerable attention recently as the first realization of a metallic, yet topologically 
nontrivial state of matter~\cite{Wan11,Ran11,Burkov11,Xu11}, as anticipated some time ago by Volovik~\cite{Volovik}. 
Observation of the closely-related Dirac semimetals~\cite{Kane12,Fang12,Fang13,Cava13,Shen13,Hasan13,Ando11,Ando14}
clearly makes the experimental realization of Weyl semimetals only a matter of time. 

The most distinctive observable {\em spectroscopic} feature of Weyl semimetals is the presence of the so-called Fermi-arc surface states~\cite{Wan11}. 
It is of great interest, however, to find similar smoking-gun features of Weyl semimetals in {\em response}, especially in transport. 
These do exist and have been described as being consequences of chiral anomaly, i.e. anomalous nonconservation of the numbers 
of Weyl fermions of distinct chiralities~\cite{Aji12,Son12,Zyuzin12,Spivak12,Grushin12,Goswami13,Qi13,Liu13,Pesin13,Pesin14,Burkov14}.
Notably, Son and Spivak~\cite{Spivak12} have proposed that in nonmagnetic inversion-asymmetric Weyl semimetals chiral anomaly leads to a 
novel kind of weak-field magnetoresistance: negative and quadratic in the magnetic field. 

However, while chiral anomaly is a well-defined concept in the context of relativistic field theory~\cite{Adler69,Jackiw69}, where 
massless fermions in unbounded momentum space possess exact chiral symmetry, violated by the anomaly, the situation is less clear in the 
condensed matter context. Even though chiral symmetry may be formally defined in a low-energy model of a Weyl semimetal, in which the band 
dispersion is approximated as being exactly linear and unbounded, no real microscopic model of Weyl semimetal actually possesses such a symmetry, 
simply because the momentum space in this case is compact, being confined to the first Brillouin zone (BZ).  
Since the chiral symmetry is not present to begin with, it is then unclear how meaningful it is to speak of its violation by chiral anomaly
and the physical consequences of this violation. 

In this paper we clarify the issues raised above. Starting from a microscopic model of Weyl semimetal~\cite{Burkov11}, which does not 
possess chiral symmetry, we demonstrate that one may, nevertheless, define a microscopic quantity, which we call axial charge density in analogy to 
the corresponding concept in relativistic field theory, and show that this quantity may be expected to be conserved or nearly conserved in the 
absence of an external magnetic field, 
when one is not too close to the phase boundaries at which the Weyl semimetal phase disappears. We then derive hydrodynamic
(diffusion) equations, which govern coupled evolution of the axial and the total charge densities
in the presence of an external magnetic field. The near conservation of the axial charge density at the microscopic level translates into 
long relaxation time at the level of hydrodynamic equations. We demonstrate that when the axial charge relaxation time is long, 
any Weyl metal indeed possesses a large
negative magnetoresistance, which is quadratic in the magnetic field, in agreement with~\cite{Spivak12}. We show, however, that this effect is in fact even more universal than suggested in~\cite{Spivak12}, and 
characterizes magnetic Weyl semimetals just as well as the inversion-asymmetric ones.  
In this sense, quadratic negative weak-field magnetoresistance may be regarded as a universal smoking-gun transport signature of Weyl semimetals 
and Weyl metals.

We start from the microscopic model of Weyl semimetal in a magnetically-doped topological (TI)  and normal (NI) insulator multilayer, introduced 
by us before~\cite{Burkov11}, which has the important virtue of being the simplest realistic model of Weyl semimetal
\beq
\label{eq:1}
\cH(\bk) = v_F \tau^z (\hat z \times \bsigma) \cdot \bk + b \sigma^z + \hat \Delta(k_z),
\eeq
where $\hat \Delta(k_z) = \Delta_S \tau^x + \frac{\Delta_D}{2} (\tau^+ e^{i k_z d} + \textrm{h.c.})$. 
$\bsigma$ and $\btau$ in Eq.~\eqref{eq:1} are Pauli matrices, describing the spin and the {\em which surface} 
pseudospin degrees of freedom, $b$ is the spin splitting due to magnetized impurities, $\Delta_{S,D}$ are 
tunnelling matrix elements, describing tunnelling between TI surface states in the same or neighboring TI layers, 
and $d$ is the superlattice period in the growth ($z$) direction. We will take $\Delta_{S,D}$ to be nonnegative for concreteness. 

Eq.~\eqref{eq:1} has a Weyl semimetal phase when $b_{c1} \leq b \leq b_{c2}$, where $b_{c1} = | \Delta_S - \Delta_D|$
and $b_{c2} = \Delta_S + \Delta_D$. The two Weyl nodes are located on the $z$-axis in momentum space at 
points $k^z_{\pm} = \pi/d \pm k_0$, where $k_0  = d^{-1} \arccos[(\Delta_S^2 + \Delta_D^2 - b^2)/2 \Delta_S \Delta_D]$. 
The Weyl nodes are interchanged by the spatial inversion transformation, with the inversion centre placed midway between 
the top and bottom surfaces of any TI or NI layer, $\cI: \cH(\bk) \rightarrow \tau^x \cH(- \bk) \tau^x$. 

We now introduce the axial charge density operator, which is analogous to the total charge density in every aspect, except changes sign 
when the chiralities of the Weyl nodes are interchanged (generalization of this concept to multiple Weyl node pairs is obvious). 
It can be defined rigorously and uniquely based on symmetry considerations. Namely, we define the axial charge density $n_a$ as a local 
operator, that is odd under inversion $\cI$ and $z \rightarrow -z$ reflections, even under time reversal, but odd under 
time reversal, combined with rotation of the spin quantization axis by $\pi$ around either $x$- or $y$-axis.  
This uniquely determines the explicit representation of the axial charge density operator to be
\beq
\label{eq:3}
\hat n_a = \tau^y \sigma^z. 
\eeq
One may easily check~\cite{Burkov12} that adding the term $- \mu_a \hat n_a$, where $\mu_a$ is the axial chemical potential, to the Hamiltonian Eq.~\eqref{eq:1}, shifts the Weyl nodes in opposite directions in energy, giving rise to the energy difference 
\beq
\label{eq:4}
\Delta \epsilon = \frac{2 \mu_a \tilde v_F}{\Delta_S d}, 
\eeq
where 
\beq
\label{eq:5}
\tilde v_F = \frac{d}{2 b} \sqrt{(b^2 - b_{c1}^2)(b_{c2}^2- b^2)}, 
\eeq
is the $z$-component of the Fermi velocity at the location of the Weyl nodes. 

We now ask the following question: does $\hat n_a$ represent a conserved quantity, as it would in a low-energy model of Weyl semimetal? To answer this we need to evaluate the commutator of $\hat n_a$ with the Hamiltonian $\cH(\bk)$. 
It is convenient at this point to apply the following canonical transformation to all the operators: $\sigma^{\pm} \rightarrow \tau^z \sigma^{\pm},\,\, \tau^{\pm} \rightarrow
\sigma^z \tau^{\pm}$~\cite{Burkov11}. Evaluating the commutator at the Weyl node locations, we now obtain
\beq
\label{eq:7}
\left[\cH(\bk), \hat n_a\right]_{k^z_{\pm}} =  i \frac{b^2 - \Delta_D^2 + \Delta_S^2}{\Delta_S}  \tau^z \sigma^z. 
\eeq
This means that $n_a$ may indeed be a conserved quantity in the Weyl semimetal or weakly-doped Weyl metal, provided $\Delta_D \geq \Delta_S$ and $b = \sqrt{b_{c1} b_{c2}}$, 
i.e. the magnitude of the spin splitting is exactly the geometric mean of its lower- and upper-critical values, at which the transitions out of the Weyl semimetal phase occur. 
Otherwise, the commutator is nonzero and $n_a$ is not conserved. However, as will be shown below, the relevant relaxation time may still be long, even when the above 
condition is not exactly satisfied, in which case the axial charge density is still a physically meaningful quantity. 

We now want to derive hydrodynamic transport equations (diffusion equations) for both the axial charge density $n_a(\br, t)$ and the total charge density $n(\br, t)$. 
As will be shown below, what is known as chiral anomaly will be manifest at the level of these hydrodynamic equations as a {\em coupling between 
$n_a$ and $n$ in the presence of an external magnetic field}. This coupling leads to significant observable magnetotransport effects, provided the 
axial charge relaxation time, calculated below, is long enough. 

To proceed with the derivation, we add a constant uniform magnetic field in the $\hat z$ direction $\bB = B \hat z$ and a scalar impurity potential $V(\br)$, 
whose precise form will be specified later. 
Adopting Landau gauge for the vector potential $\bA = x B \hat y$, the second-quantized Hamiltonian of our system may be written as
\beqa
\label{eq:8}
&&H = \sum_{n a k_y k_z} \epsilon_{n a}(k_z) c^\dg_{n a k_y k_z} c^\pdg_{n a k_y k_z} \nonumber \\
&+& \sum_{n a k_y k_z, n'  a'  k_y' k_z'} \langle n, a, k_y, k_z | V | n', a', k_y', k_z' \rangle c^\dg_{n a k_y k_z} c^\pdg_{n' a' k_y' k_z'}. \nonumber \\
\eeqa 
Here $\epsilon_{n a}(k_z)$ are Landau-level (LL) eigenstate energies of a clean multilayer in magnetic field, $n=0,1,2,\ldots$ is the main LL
index, $k_y$ is the Landau-gauge intra-LL orbital quantum number, $k_z$ is the conserved component of the crystal momentum 
along the $z$-direction, and $a = (s, t)$ is a composite index (introduced for 
compactness of notation), consisting of $s = \pm$, which labels the electron- ($s = +$) and hole- ($s = -$) like sets of Landau levels, and $t = \pm$, which 
labels the two components of a Kramers doublet of LLs, degenerate at $b = 0$.
Explicitly we have 
\beq
\label{eq:9}
\epsilon_{n a}(k_z) = s \sqrt{2 \omega_B^2 n + m_t^2(k_z)} \equiv s \epsilon_{n t}(k_z), 
\eeq
where $\omega_B = v_F/ \ell_B$ is the Dirac cyclotron frequency and $\ell_B = 1/\sqrt{e B}$ is the magnetic length. 
We will use units in which $\hbar = c = 1$ throughout. 
The ``Dirac massess" $m_t(k_z)$ are given by $m_t(k_z) = b + t \Delta(k_z)$ where $ \pm \Delta(k_z) = \pm \sqrt{\Delta_S^2 + \Delta_D^2 + 2 \Delta_S \Delta_D \cos(k_z d)}$ 
are the two eigenvalues of the $\hat \Delta(k_z)$ operator. 

The LL eigenstates have the following form, typical for LLs in Dirac systems
\beqa
\label{eq:10}
| n, a, k_y, k_z \rangle&=& \sum_{\tau} \left[z^a_{n \upa \tau}(k_z) | n -1, k_y, k_z, \upa, \tau \rangle\right. \nonumber \\
&+&\left. z^a_{n \da \tau}(k_z) | n, k_y, k_z, \da, \tau \rangle \right]. 
\eeqa 
Here 
\beq
\label{eq:11}
\langle \br | n, k_y, k_z, \sigma, \tau \rangle = \frac{1}{\sqrt{L_z}} e^{i k_z z} \phi_{n k_y}(\br) | \sigma, \tau \rangle, 
\eeq
$\phi_{n k_y}(\br)$ are the Landau-gauge orbital wavefunctions, and $\sigma, \tau$ are the spin and pseudospin indices 
respectively. Finally, the four-component eigenvector $| z^a_{n}(k_z) \rangle$ may be written as a tensor product of the two-component spin and pseudospin eigenvectors, 
i.e. $| z^a_{n}(k_z) \rangle = | v^a_{n}(k_z) \rangle \otimes | u^a(k_z) \rangle$, where 
\beqa
\label{eq:12}
&&|v^{s t}_{n}(k_z) \rangle = \frac{1}{\sqrt{2}} \left(\sqrt{1 + s \frac{m_t(k_z)}{\epsilon_{n t}(k_z)}}, - i s \sqrt{1 - s \frac{m_t(k_z)}{\epsilon_{n t}(k_z)}} \right), \nonumber \\
&&|u^t(k_z) \rangle = \frac{1}{\sqrt{2}} \left(1, t \frac{\Delta_S + \Delta_D e^{- i k_z d}}{\Delta(k_z)} \right). 
\eeqa
As in all Dirac systems, the lowest $n = 0$ LL is special and needs to be considered separately. The $s$ quantum number is absent in this case and  
taking $B > 0$ for concreteness, we have $\epsilon_{n t}(k_z) = - m_t(k_z)$, and $|v^t_{0}(k_z) \rangle = (0,1)$. 

To proceed, we will make the standard assumption that the impurity potential obeys Gaussian distribution, with $\langle V(\br) V(\br') \rangle = \gamma^2 \delta(\br - \br ')$.
To simplify calculations further we will also assume that the momentum transfer due to the impurity scattering is smaller than the size of the BZ, i.e. $| k_z - k_z' | d \ll 1$. 
In this case $\langle u^t(k_z)| u^{t'}(k_z') \rangle \approx \delta_{t t'}$, i.e. the $t$ quantum number may be assumed to be approximately preserved during the impurity scattering. 

We treat the impurity scattering in the standard self-consistent Born approximation (SCBA).  
The retarded SCBA self-energy satisfies the equation
\beqa
\label{eq:13}
\Sigma^R_{n a k_y k_z }(\omega)&=&\frac{1}{L_z} \sum_{n' a' k_y'  k_z'} \langle |\langle n, a, k_y, k_z| V |n', a', k_y', k_z'\rangle|^2 \rangle \nonumber \\
&\times& G^R_{n' a' k_y' k_z'}(\omega), 
\eeqa
We will assume that the Fermi energy $\epsilon_F$ is positive, i.e. the Weyl semimetal is electron-doped, and large enough that the impurity-scattering-induced broadening of the density of states is small on the scale of the Fermi energy $\epsilon_F$~\cite{footnote1}.  We can then restrict ourselves to the electron-like states with $s=+$ (we will drop the $s$ index henceforth for brevity), and easily solve the SCBA equation analytically. We obtain
\beq
\label{eq:14}
\textrm{Im} \Sigma^R_{n t k_z} \equiv -\frac{1}{2 \tau_t(k_z)} = - \frac{1}{2 \tau} \left[1 + \frac{m_t(k_z) \langle m_t \rangle}{\epsilon_F^2} \right], 
\eeq
where $1/ \tau = \pi \gamma^2 g(\epsilon_F)$ and 
\beq
\label{eq:15}
g(\epsilon_F) = \frac{1}{2 \pi \ell_B^2} \int_{-\pi/d}^{\pi/d} \frac{d k_z}{2 \pi} \sum_{n t} \delta[\epsilon_{n t}(k_z) - \epsilon_F],
\eeq
is the density of states at Fermi energy. 
We have also introduced the Fermi-surface average of $m_t(k_z)$ as
\beq
\label{eq:16}
\langle m_t \rangle = \frac{1}{2 \pi \ell_B^2 g(\epsilon_F)} \int_{-\pi/d}^{\pi/d} \frac{d k_z}{2 \pi} \sum_{n t} m_t(k_z) \delta[\epsilon_{n t}(k_z) - \epsilon_F]. 
\eeq

All the necessary information about the density response of our system is contained in the diffusion propagator, or diffuson $\cD$, given 
by the sum of ladder impurity-averaging diagrams~\cite{Altland}. 
This is evaluated in the standard manner and we obtain
\beq
\label{eq:17}
\cD^{-1} (\bq, \Omega)= 1 - I(\bq, \Omega), 
\eeq
where $I$ is a $16 \times 16$ matrix, given by
\beqa
\label{eq:18}
&&I_{\alpha_1 \alpha_2, \alpha_3 \alpha_4}(\bq, \Omega) = \frac{\gamma^2}{L_x L_y L_z} \int d^3 r d^3 r' e^{-i \bq \cdot (\br - \br')} \nonumber \\
&\times&G^R_{\alpha_1 \alpha_3}(\br, \br'| \Omega) G^A_{\alpha_4 \alpha_2}(\br', \br | 0), 
\eeqa
where we have introduced a composite index $\alpha = (\sigma, \tau)$ to simplify the notation. 
The impurity-averaged Green's functions $G^{R,A}$ are given by
\beq
\label{eq:19}
G^{R,A}_{\alpha \alpha'}(\br, \br' |\Omega) = \sum_{n t k_y k_z} \frac{\langle \br, \alpha | n,t,k_y,k_z \rangle \langle n,t,k_y,k_z | \br', \alpha' \rangle}
{\Omega - \xi_{n t}(k_z) \pm i / 2 \tau_t(k_z)}, 
\eeq
where $\xi_{n t}(k_z) = \epsilon_{n t}(k_z) - \epsilon_F$. 

In general, the evaluation of Eq.~\eqref{eq:18} is a rather complicated task, primarily due to the fact that the impurity scattering will mix different LLs. 
At this point we will thus specialize to the case of transport along the $z$-direction only, as this is where we can expect 
chiral anomaly to be manifest. In this case the contributions of different LLs to Eq.~\eqref{eq:18} decouple. 
Setting $\bq = q \hat z$, we obtain
\beqa
\label{eq:20}
&&I_{\alpha_1 \alpha_2, \alpha_3 \alpha_4} (q, \Omega)  = \frac{\gamma^2}{2 \pi \ell_B^2 L_z} \nonumber \\
&\times& \sum_{n t t' k_z} \frac{\langle \alpha_1| z^{t}_{n}(k_z + q/2) \rangle \langle z^{t}_{n}(k_z + q/2) | \alpha_3 \rangle}
{\Omega - \xi_{n t}(k_z + q/2) + i/ 2 \tau_t(k_z + q/2)} \nonumber \\
&\times&\frac{\langle \alpha_4 | z^{t'}_{n}(k_z - q/2) \rangle \langle z^{t'}_{n}(k_z - q/2) | \alpha_2 \rangle}{-\xi_{n t'}(k_z - q/2) - i/ 2 \tau_t'(k_z - q/2)}. 
\eeqa

As mentioned above, $I$ and $\cD^{-1}$ are large $16 \times 16$ matrices, which contain a lot of information of no interest to us. 
We are interested only in hydrodynamic physical quantities, with long relaxation times. All such quantities need to be identified, 
if they are expected to be coupled to each other. 
One such quantity is obviously the total charge density $n(\br, t)$, which has an infinite relaxation time due to the exact conservation of particle number. 
Another is the axial charge density $n_a(\br, t)$, which, as discussed above, may be almost conserved under certain conditions. 
On physical grounds, we expect no other hydrodynamic quantities to be present in our case. We are thus only interested in the $2 \times 2$ block of the matrix $\cD^{-1}$, which corresponds to the coupled evolution of the total and the axial charge densities. 
To separate out this block, we apply the following transformation to the inverse diffuson matrix
\beq
\label{eq:21}
\cD^{-1}_{a_1 b_1,  a_2 b_2} = \frac{1}{2} (\sigma^{a_1} \tau^{b_1})_{\alpha_2 \alpha_1} \cD^{-1}_{\alpha_1 \alpha_2, \alpha_3 \alpha_4}
(\sigma^{a_2} \tau^{b_2})_{\alpha_3 \alpha_4}, 
\eeq
where $a_{1,2}, b_{1,2} = 0, x, y, z$. The components of interest to us are $a_{1,2} = b_{1,2} = 0$ which corresponds to the total charge density, $a_{1,2} = 0, b_{1,2} = y$, 
which corresponds to the axial charge density, and the corresponding cross-terms.   

We will be interested in the hydrodynamic regime, which corresponds to low frequencies and long wavelengths, i.e. $\Omega \tau \ll 1$ and 
$v_F q \tau \ll 1$. We will also assume that the magnetic field is weak, so that $\omega_B \ll \epsilon_F$. 
Finally, we will assume that the Fermi energy is close enough to the Weyl nodes, so that only the $t = -$ states participate in transport and 
$\langle m_- \rangle \approx 0$, since $m_-(k_z)$ changes sign at the nodes~\cite{Burkov14}. 

In accordance with the above assumptions, we expand the inverse diffusion propagator to leading order in $\Omega \tau$, $v_F q \tau$ and $\omega_B/ \epsilon_F$ 
and obtain after a straightforward but lengthy calculation 
\beqa
\label{eq:22}
\cD^{-1}(q, \Omega) = \left(
\begin{array}{cc}
 -i \Omega \tau + D q^2 \tau & -i q \Gamma \tau \\
 - i q \Gamma \tau & -i \Omega \tau + D q^2 \tau + \tau/ \tau_a
 \end{array}
 \right). \nonumber \\
 \eeqa 
Here $D = \tilde v_F^2 \tau \langle m_-^2 \rangle/ \epsilon_F^2$ is the charge diffusion constant, associated with the diffusion in the $z$-direction, 
$\Gamma = e B / 2 \pi^2 g(\epsilon_F)$ is the total charge-axial charge coupling coefficient and 
\beq
\label{eq:23}
\frac{1}{\tau_a} = \frac{1 - (\tilde v_F/ \Delta_S d)^2}{(\tilde v_F/ \Delta_S d)^2 \tau}, 
\eeq
is the axial charge relaxation rate. 
Several comments are in order here. First, note that the axial charge relaxation rate $1/\tau_a \geq 0$, as it should be, and vanishes when $\tilde v_F = \Delta_S d$. 
It is easy to see that this is identical to the condition of the vanishing of the commutator of the axial charge operator with the Hamiltonian Eq.~\eqref{eq:7}.
Henceforth we will assume that this condition is nearly satisfied so that $\tau_a \gg \tau$.
Second, the situation when $\tilde v_F = \Delta_S d$ and thus $1/\tau_a$ appears to vanish, actually needs to be treated with some care. Namely, 
the condition $\tilde v_F = \Delta_S d$ may be satisfied exactly only in the limit $\epsilon_F \rightarrow 0$. The Fermi velocity  depends on the Fermi energy as~\cite{Pesin14}
\beq 
\label{eq:23}
\tilde v_F(\epsilon_F) = \frac{d}{2 (b + \epsilon_F)} \sqrt{[(b + \epsilon_F)^2 - b_{c1}^2] [b_{c2}^2 - (b + \epsilon_F)^2]}. 
\eeq
When $b = \sqrt{b_{c1} b_{c2}}$ and thus $\tilde v_F(0) = \Delta_S d$, the Fermi energy dependence of $\tilde v_F$ needs to be taken into account. 
Expanding to leading non vanishing order in $\epsilon_F$ we obtain in this case
\beq
\label{eq:24}
\frac{1}{\tau_a} = \frac{\epsilon_F^2}{\Delta_S^2 \tau}, 
\eeq
i.e. $1/\tau_a$ is in fact always finite, but may be very small. We can estimate the minimal value of the axial charge relaxation rate by 
setting $\epsilon_F \approx 1/\tau$ in Eq. ~\eqref{eq:24}, which gives $(\tau/\tau_a)_{min} \approx 1/ (\Delta_S \tau)^2$. 

We may now write down the coupled diffusion equations for the total and axial charge densities, which correspond to the propagator Eq.~\eqref{eq:22}. 
These equations read
\beqa
\label{eq:25}
\frac{\partial n}{\partial t}&=&D \frac{\partial^2 n}{\partial z^2} + \Gamma \frac{\partial n_a}{\partial z}, \nonumber \\
\frac{\partial n_a}{\partial t}&=&D \frac{\partial^2 n_a}{\partial z^2} - \frac{n_a}{\tau_a} + \Gamma \frac{\partial n}{\partial z}. 
\eeqa
Eq.~\eqref{eq:25} is our main result. 
Manifestation of chiral anomaly in these equations is the coupling between the total and the axial charge densities, proportional to the 
applied magnetic field. 
Since the total particle number is conserved, the right-hand side of the first of Eqs.~\eqref{eq:25} must be equal to minus the divergence of the total particle current. 
Then we obtain the following expression for the density of the charge current in the $z$-direction
\beq
\label{eq:26}
j = - \frac{\sigma_0}{e} \frac{\partial \mu}{\partial z} - \frac{e^2 B}{2 \pi^2} \mu_a, 
\eeq
where $\sigma_0 = e^2 g(\epsilon_F) D$ is the zero-field diagonal charge conductivity, $\mu$ and $\mu_a$ are the total and axial electrochemical potentials 
and we have used $\delta n = g(\epsilon_F) \delta \mu$, $\delta n_a = g(\epsilon_F) \delta \mu_a$. The last relation is valid when $\tilde v_F/ \Delta_S d$ is close
to unity, as seen from Eq.~\eqref{eq:4}. 
Thus chiral anomaly manifests in an extra contribution to the charge current density, proportional to the magnetic field and the axial electrochemical potential. 
This is known as chiral magnetic effect (CME) in the literature~\cite{Kharzeev,Franz13,Burkov13}. 
Note that the CME contribution to the current exists only away from equilibrium~\cite{Franz13,Burkov13}, i.e. the second term in Eq.~\eqref{eq:26} should never be interpreted as an equilibrium current, driven by a static magnetic field~\cite{Scalapino93}. 

To find measurable consequences of the CME contribution to the charge current, we consider a steady-state situation, with a fixed current density $j$ flowing through the 
sample in the $z$-direction. We want to find the corresponding electrochemical potential drop and thus the conductivity. 
Assuming the current density is uniform, we obtain from the second of Eqs.~\eqref{eq:24}
\beq
\label{eq:27}
n_a = \Gamma \tau_a \frac{\partial n}{\partial z}, 
\eeq 
which is the nonequilibrium axial charge density, induced by the current and the corresponding electrochemical potential gradient. 
Substituting this into the expression for the charge current density Eq.~\eqref{eq:26}, we finally obtain the following result
for the conductivity
\beq
\label{eq:28}
\sigma = \sigma_0 + \frac{e^4 B^2 \tau_a}{4 \pi^4 g(\epsilon_F)}. 
\eeq
In the limit when $\epsilon_F$ is not far from the Weyl nodes, such that the dispersion may be assumed to be linear, 
we have $g(\epsilon_F) = \epsilon_F^2/ \pi^2 v_F^2 \tilde v_F$, which gives
\beq
\label{eq:29}
\Delta \sigma = \sigma - \sigma_0 = \frac{e^2 \tilde v_F \tau_a}{(2 \pi v_F)^2} \left(\frac{e^2 v_F^2 B}{\epsilon_F} \right)^2,
\eeq
which agrees with the Son and Spivak result~\cite{Spivak12}. 
Thus we see that a measurable consequence of CME is a positive magnetoconductivity, proportional to $B^2$ in the 
limit of a weak magnetic field. 
This of course needs to be compared with the classical negative magnetoconductivity, which is always 
present and arises from the $B^2$ corrections to the diffusion constant $D$, which we have neglected
\beq
\label{eq:30}
\frac{\Delta \sigma_{c \ell}}{\sigma_0} \sim - (\omega_c \tau)^2, 
\eeq 
where $\omega_c = e v_F^2 B/ \epsilon_F$ is the cyclotron frequency. 
This gives 
\beq
\label{eq:31}
\left|\frac{\Delta \sigma}{\Delta \sigma_{c \ell}}\right| \sim \frac{\tau_a/\tau}{(\epsilon_F \tau)^2}. 
\eeq
Thus the CME-related positive magnetoconductivity will dominate the classical negative magnetoconductivity, provided 
$\tau_a$ is long enough. 

As a final comment we note that we have so far ignored the Zeeman effect due to the applied magnetic field. 
Its effect is to modify the spin-splitting parameter $b$ as $b \rightarrow b + g \mu_B B/2$. 
In principle, the dependence on $b$ does enter into our final results through the dependence 
of the Fermi velocity $\tilde v_F$ on $b$. Naively, this will then generate an additional {\em linear} magnetoconductivity, which 
may be expected to dominate the quadratic one at small fields.
However, the condition of large $\tau_a$, which is the same as $\tilde v_F/ \Delta_S d \approx 1$, is equivalent to 
the condition $b_{c1} \ll b \ll b_{c2}$, in which case the dependence of $\tilde v_F$ on $b$ becomes negligible. 
Thus, in the regime in which the positive magnetoconductivity dominates the negative classical one, and is thus observable, 
one may also expect a negligible linear magnetoconductivity in any type of Weyl metal. 
 
\begin{acknowledgments}
Financial support was provided by NSERC of Canada. 
\end{acknowledgments}


\begin{thebibliography}{99}
\bibitem{Wan11} X. Wan, A.M. Turner, A. Vishwanath, and S.Y. Savrasov, Phys. Rev. B {\bf 83}, 205101 (2011). 
\bibitem{Ran11} K.-Y. Yang, Y.-M. Lu, and Y. Ran, Phys. Rev. B {\bf 84}, 075129 (2011).
\bibitem{Burkov11} A.A. Burkov and L. Balents, Phys. Rev. Lett. {\bf 107}, 127205 (2011).
\bibitem{Xu11} G. Xu, H.-M. Weng, Z.-J. Wang, X. Dai, and Z. Fang, Phys. Rev. Lett. {\bf 107}, 186806 (2011).
\bibitem{Volovik} G.E. Volovik, {\em The Universe in a Helium Droplet} (Clarendon Press, Oxford, 2003); Lect. Notes Phys. {\bf 718}, 31 (2007).
\bibitem{Kane12} S.M. Young, S. Zaheer, J.C.Y. Teo, C.L. Kane, and E.J. Mele, Phys. Rev. Lett. {\bf 108}, 140405 (2012).
\bibitem{Fang12} Z.-J. Wang, Y. Sun, X.-Q. Chen, C. Franchini, G. Xu, H.-M. Weng, X. Dai, and Z. Fang, Phys. Rev. B {\bf 85}, 195320 (2012).
\bibitem{Fang13} Z.J. Wang, H.-M. Weng, Q.-S. Wu, X. Dai, and Z. Fang, Phys. Rev. B {\bf 88}, 125427 (2013).
\bibitem{Cava13} S. Borisenko, Q. Gibson, D. Evtushinsky, V. Zabolotnyy, B. B\"uchner, and R.J. Cava, Phys. Rev. Lett. {\bf 113}, 027603 (2014). 
\bibitem{Shen13} Z.K. Liu, B. Zhou, Z.J. Wang, H.M. Weng, D. Prabhakaran, S.-K. Mo, Y. Zhang, Z.X. Shen, Z. Fang, X. Dai, Z. Hussain, and Y.L. Chen,
Science {\bf 343}, 864 (2014). 
\bibitem{Hasan13} M. Neupane, S.-Y. Xu, R. Sankar, N. Alidoust, G. Bian, C. Liu, I. Belopolski, T.-R. Chang, H.-T. Jeng, H. Lin, A. Bansil, F. Chou,
and M.Z. Hasan, Nature Commun. {\bf 5}, 3786 (2014). 
\bibitem{Ando11} T. Sato, K. Segawa, K. Kosaka, S. Souma, K. Nakayama, K. Eto, T. Minami, Y. Ando, and T. Takahashi, Nat. Phys. {\bf 7}, 840 (2011). 
\bibitem{Ando14} M. Novak, S. Sasaki, K. Segawa, and Y. Ando, arXiv:1408.2183 (unpublished)
\bibitem{Aji12} V. Aji, Phys. Rev. B {\bf 85}, 241101 (2012).
\bibitem{Son12} D.T. Son and N. Yamamoto, Phys. Rev. Lett. {\bf 109}, 181602 (2012).
\bibitem{Zyuzin12} A.A. Zyuzin and A.A. Burkov, Phys. Rev. B {\bf 86}, 115133 (2012).
\bibitem{Spivak12} D.T. Son and B.Z. Spivak, Phys. Rev. B {\bf 88}, 104412 (2013).
\bibitem{Grushin12} A.G. Grushin, Phys. Rev. D {\bf 86}, 045001 (2012).
\bibitem{Goswami13} P. Goswami and S. Tewari, Phys Rev. B {\bf 88}, 245107 (2013).
\bibitem{Qi13} P. Hosur and X.-L. Qi, C. R. Physique {\bf 14}, 857 (2013).
\bibitem{Liu13} C.-X. Liu, P. Ye, X.-L. Qi, Phys. Rev. B {\bf 87}, 235306 (2013).
\bibitem{Pesin13}  S.A. Parameswaran, T. Grover, D.A. Abanin, D.A. Pesin, and A. Vishwanath, Phys. Rev. X {\bf 4}, 031035 (2014). 
\bibitem{Pesin14} I. Panfilov, A.A. Burkov, and D.A. Pesin, Phys. Rev. B {\bf 89}, 245103 (2014). 
\bibitem{Burkov14} A.A. Burkov, Phys. Rev. Lett. {\bf 113}, 187202 (2014). 
\bibitem{Adler69} S. Adler, Phys. Rev. {\bf 177}, 2426 (1969).
\bibitem{Jackiw69} J.S. Bell and R. Jackiw, Nuovo Cimento {\bf 60A}, 4 (1969).
\bibitem{Burkov12} A.A. Zyuzin, S. Wu, and A.A. Burkov, Phys. Rev. B {\bf 85}, 165110 (2012). 
\bibitem{footnote1} For a treatment of transport in Weyl semimetals in the limit $\epsilon_F \rightarrow 0$ see
R.R. Biswas and S. Ryu, Phys. Rev. B {\bf 89}, 014205 (2014). 
\bibitem{Altland} See e.g. A. Altland and B. Simons, {\em Condensed Matter Field Theory}, (Cambridge University Press, Cambridge, 2010).  
\bibitem{Kharzeev} K. Fukushima, D.E. Kharzeev, and H.J. Warringa, Phys. Rev. D {\bf 78}, 074033 (2008).
\bibitem{Franz13} M.M. Vazifeh and M. Franz, Phys. Rev. Lett. {\bf 111}, 027201 (2013). 
\bibitem{Burkov13} Y. Chen, S. Wu, and A.A. Burkov, Phys. Rev. B {\bf 88}, 125105 (2013). 
\bibitem{Scalapino93} For a general discussion of the difference between equilibrium and nonequilibrium response see, e.g., D.J. Scalapino, S.R. White, and S.-C. Zhang, Phys. Rev. B {\bf 47}, 7995 (1993). 
\end{thebibliography}
\end{document}